\documentclass[a4paper]{jpconf}
\usepackage{graphicx}

\usepackage{amsmath}
\usepackage{braket}

\begin{document}
\title{Study of $K_S$ semileptonic decays and $\mathcal{CPT}$ test with the KLOE detector}

\author{Daria Kami{\'n}ska \\ on behalf of the KLOE-2 Collaboration}

\address{The Marian Smoluchowski Institute of Physics, Jagiellonian University \\ 
{\L}ojasiewicza 11, 30-348 Krak{\'o}w, Poland}

\ead{daria.kaminska@doctoral.uj.edu.pl}


\begin{abstract}
Study of semileptonic decays of neutral kaons allows to perform a test of discrete symmetries, as
well as basic principles of the Standard Model. In this paper a general review on dependency
between charge asymmetry constructed for semileptonic decays of short- and long-lived kaons and
$\mathcal{CPT}$ symmetry is given.
The current status of determination of charge asymmetry for short-lived kaon, obtained by
reconstruction of about $10^5$ $K_S \rightarrow \pi e \nu$ decays collected at DA$\Phi$NE with the KLOE detector is also
reviewed.
\end{abstract}

\section{Introduction}
	Discrete symmetries of nature  such as charge conjugation ($\mathcal{C}$), parity	($\mathcal{P}$)
	or time reversal ($\mathcal{T}$) do not lead to new conserved quantities. Nevertheless,
	all	of the mentioned symmetries plays an important role in particle physics, especially
	in	calculations of the cross sections and decay rates.
	Weak	interaction does not conserve the $\mathcal{C}$, $\mathcal{P}$, $\mathcal{T}$ or combined
	$\mathcal{CP}$ symmetry. However, up to now, there is no
	indication	of $\mathcal{CPT}$ symmetry violation~\cite{cpt_table}, which would also imply the break of
	Lorentz	symmetry~\cite{cpt_lorentz}. A special role in $\mathcal{CPT}$ violation searches plays a neutral kaon
	system	which, due to a sensitivity to a variety of symmetry violation effects, is one
	of	the best candidates for such kind of studies. One of the possible tests is based on
	comparison	between semileptonic asymmetry in $K_S$ decays ($A_S$) and the analogous
	asymmetry	in $K_L$ decays ($A_L$).

\section{Test of $\mathcal{CPT}$ symmetry violation through semileptonic decays in neutral kaon system}
\subsection{Semileptonic decays in neutral kaon system}
	Neutral	kaons are the lightest particles which contain a strange quark. Observed	short-lived $K_S$
	and long-lived $K_L$ are linear combinations of strange 
	eigenstates	($K^0$ and $\bar{K^0}$):
 \begin{equation}
 \begin{aligned}
  \ket{ K_{L} } = \frac{1}{\sqrt{2(1+|\epsilon_{L}|^2)}  } \left(  (1+ \epsilon_{L})  \ket{K^0}
  -(1- \epsilon_{L})  \ket{\bar{K^0}} \right), \\
   \ket{ K_{S} } = \frac{1}{\sqrt{ 2(1+|\epsilon_{S}|^2) }} \left( (1+ \epsilon_{S})  \ket{K^0}  +
  (1- \epsilon_{S})  \ket{\bar{K^0}} \right),   
  \label{ks_kl}
  \end{aligned}
  \end{equation}
 where introduced small parameters $\epsilon_{L}$ and
	$\epsilon_{S}$ can be rewritten to separate $\mathcal{CP}$ and $\mathcal{CPT}$ violation
	parameters 		$\epsilon_{K}$ and $\delta_{K}$, respectively: 
  \begin{align}
    \begin{aligned}
     \epsilon_{S} = \epsilon_{K} + \delta_{K}, \\
     \epsilon_{L} = \epsilon_{K} - \delta_{K}.  
    \end{aligned}
    \label{epsilon_and_symmetries}
    \end{align}
 
\begin{figure}[h!]
		\centering
		\includegraphics[width=0.45\textwidth]{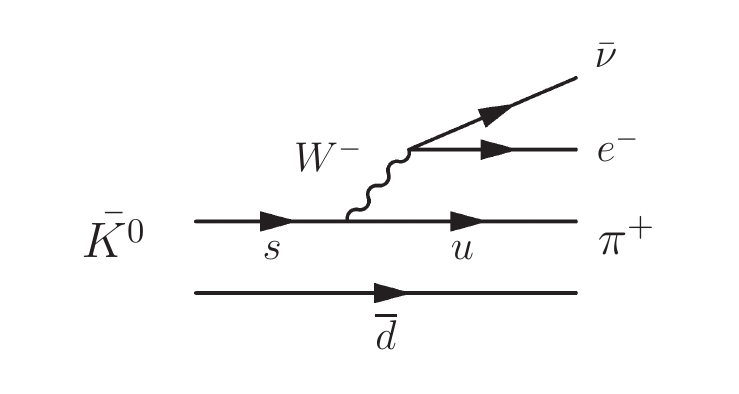}
		\includegraphics[width=0.45\textwidth]{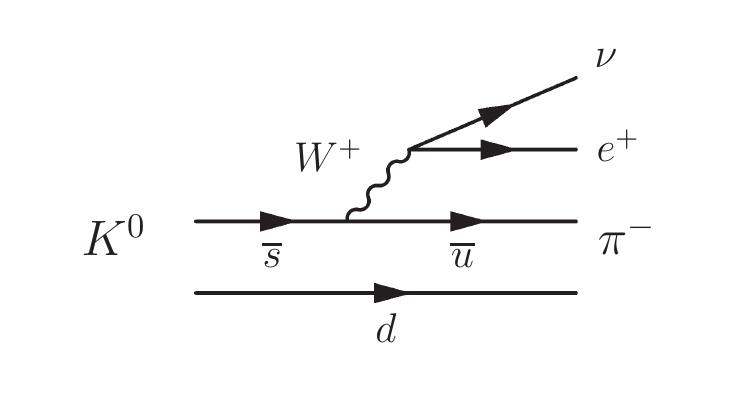}
		\caption{Feynman diagrams for $K^0$ and $\bar{K^0}$ semileptonic decay.}
		\label{feynmann_diagrams_semi}
	\end{figure}

 In this paper a particular emphasis will be given to semileptonic decays of K-short and K-long states
	($K_{S/L} \rightarrow \pi e \nu$). According to Eq.~\ref{ks_kl}, only the four
	possible decays of strange eigenstates should be considered:
		\begin{align}
		\begin{aligned}
			K^0 \rightarrow \pi^- e^+ \nu, & ~~ & 
			K^0 \rightarrow \pi^+ e^- \bar{\nu}, \\
			\bar{K^0} \rightarrow \pi^+ e^- \bar{\nu}, & ~~ &
			\bar{K^0} \rightarrow \pi^- e^+ \nu. 
		\end{aligned}
		\end{align}
	However, in the Standard Model the decay of $K^0$ (or $\bar{K^0}$) state is associated with the
 transition of the $\bar{s}$ quark into $\bar{u}$ quark (or $s$ into $u$) and emission of the
 charged boson. 
 Change of strangeness ($\Delta S$) implies the corresponding
 change of electric charge  ($\Delta Q$) (see Figure~\ref{feynmann_diagrams_semi}). This is so called $\Delta S = \Delta Q$ rule.
 Therefore, decays  of $K^0\rightarrow \pi^- e^+ \nu$ and $\bar{K^0} \rightarrow \pi^+ e^-
 \bar{\nu}$ are present but $ K^0 \rightarrow \pi^+ e^-\bar{\nu}$ and $\bar{K^0} \rightarrow \pi^- e^+
 \nu$ are forbidden by if $\Delta Q = \Delta S$.

 Decay amplitudes for semileptonic decays of states $\ket{ K^0}$ and $\ket{ \bar{K^0}}$  can
 be written as~follows~\cite{handbook_cp}:					
   \begin{equation}
    \begin{aligned}
    \label{aplitudes_def}
    \bra{\pi^{-} e^{+} \nu } H_{weak}  \ket{K^{0}}              &  = \mathcal{A_+} ,   &    
    \bra{\pi^{+} e^{-} \bar{\nu} } H_{weak}  \ket{ \bar{K^{0}}} &  = \mathcal{\bar{A}_{-}},\\
    \bra{\pi^{+} e^{-} \bar{\nu} } H_{weak}  \ket{K^{0}}        &  = \mathcal{A_{-}},   
				&
    \bra{\pi^{-} e^{+} \nu } H_{weak}  \ket{ \bar{K^{0}}}       &  =  \mathcal{\bar{A}_{+}},
    \end{aligned}
   \end{equation} 
 where the $H_{weak}$ is the term of Hamiltonian corresponding to the weak interaction and 
 $\mathcal{A_+}, \mathcal{\bar{A}_{-}},  \mathcal{A_{-}}, \mathcal{\bar{A}_{+}}$  
 parametrize the semileptonic decay amplitudes.

	It is useful to introduce the following notation:
	\begin{align}
  \begin{aligned}
  \label{zabawne_wspolczynniki}
   x  &= \frac{\mathcal{\bar{A}_+}}{\mathcal{A}_+},   
   \ \ \ \ \ \   \bar{x} = \left( \frac{ \mathcal{A_{-}}}{\mathcal{\bar{A}_{-}}} \right)^*, 
			\ \ \ \ \ \
			y = \frac{ \mathcal{ \bar{A}_{-}^{*} } - \mathcal{A}_{+}  }{  \mathcal{ \bar{A}_{-}^{*} } +
   \mathcal{A}_{+} }, \\   
    x_{\pm}  &= \frac{ x \pm \bar{x}^{*} }{2} = \frac{1}{2}  \left[ \frac{\mathcal{\bar{A}}_+}{\mathcal{A}_+}  \pm 
    \left( \frac{\mathcal{A}_-}{\mathcal{\bar{A}}_-} \right)^* \right]. 
  \end{aligned}
 \end{align}
 For further considerations, rules for applying symmetry operators to  amplitudes of
 two spin zero systems $A$ and $B$
 (and corresponding anti-systems $\bar{A}$ and $\bar{B}$)
 could be summarized~as:
 \begin{align}
  \label{symmetries_operators}
    \bra{ \mathcal{T} \textbf{B} }  \mathcal{T} H_{wk} \mathcal{T}^{-1} \ket{\mathcal{T}
    \textbf{A} }  &= (\bra{ \textbf{B}}
    \mathcal{T} H_{wk} \mathcal{T}^{-1} \ket{\textbf{A}})^*  \nonumber \\     
    \bra{ \mathcal{CP} \textbf{B}}  \mathcal{CP} H_{wk} \mathcal{CP}^{-1} \ket{\mathcal{CP}
    \textbf{A}} & = \bra{
    \bar{ \textbf{B}} } \mathcal{CP} H_{wk} \mathcal{CP}^{-1} \ket{ 
     \bar{ \textbf{A} }}   \\     
    \bra{ \mathcal{CPT} \textbf{B}}  \mathcal{CPT} H_{wk} \mathcal{CPT}^{-1} \ket{\mathcal{CPT}
    \textbf{A}} & =
    ( \bra{ \bar{\textbf{B}} } \mathcal{CPT} H_{wk} \mathcal{CPT}^{-1} \ket{
    \bar{\textbf{A} }})^*  \nonumber
 \end{align}
  
 One obtains the relation between the semileptonic amplitudes and conservation of a particular
 symmetry by applying the presented above rules to the states presented in Eq.~\ref{aplitudes_def}. 
 These considerations are summarized in Table~\ref{table_fun1}.
	
	\begin{table}[h!]
  \centering
		\caption{Relations between discrete symmetries and semiletponic amplitudes}
  \label{table_fun1}
  \begin{tabular}{|c|c|} \hline
   Conserved quantity & Required relation \\ \hline \hline
   $\Delta S = \Delta Q$ rule & $x = \bar{x} = 0$ \\
   $\mathcal{CPT}$ symmetry & $x = \bar{x}^*$, $y=0$   \\
   $\mathcal{CP}$  symmetry & $x = \bar{x}$, $y = Im(y)$ \\
   $\mathcal{T}$   symmetry & $y= Re (y)$ \\ \hline
  \end{tabular}
 \end{table}

\subsection{Charge asymmetry}
	Charge asymmetry can be defined for semileptonic decays of $K_S$ and $K_L$ mesons in the following~way:
   \begin{equation}
    \begin{aligned}
     A_{S,L} & = 
        \frac{\Gamma(K_{S,L} \rightarrow \pi^{-} e^{+} \nu) - \Gamma(K_{S,L}
        \rightarrow
        \pi^{+} e^{-}
        \bar{\nu})}{\Gamma(K_{S,L} \rightarrow \pi^{-} e^{+} \nu) + \Gamma(K_{S,L}
        \rightarrow
        \pi^{+} e^{-} \bar{\nu})}    \\
    \end{aligned}
  \end{equation}
	and 	can be rewritten in terms of parameters introduced in Eq.~\ref{zabawne_wspolczynniki}:
   \begin{equation}
    \begin{aligned}
		     A_{S,L}   & = 2 \left[  Re\left( \epsilon_{K}\right) \pm Re \left(\delta_{K} \right) - Re (y) \pm Re( x_{-}) \right].       
    \end{aligned}
   \end{equation}

	In that case, sum and difference of the $A_{S}$ and $A_{L}$ allow to search for the $\mathcal{CPT}$
 symmetry violation, either in 
 the decay amplitudes through the parameter $y$ 
 or in the mass matrix through the parameter $\delta_{K}$: 
  \begin{equation}
   \begin{aligned}
    A_{S} + A_{L} &= 4 Re( \epsilon ) - 4 Re \left( y \right), \\
    A_{S} - A_{L}  
                   &= 4 Re( \delta_{K}) + 4 Re \left( x_{-} \right).
   \label{cpt_asymetria}
   \end{aligned}
   \end{equation}

\subsection{Experimental verification}	
	The measurement based on $1.9$ millions $K_{L} \rightarrow \pi
 e \nu$ decays produced in collisions of proton beam with a BeO target performed by KTeV
	Collaboration allowed to determine $A_L$ values~\cite{ktev_kl_charge_asymm}:
	  \begin{equation}
    A_{L} = (3.322 \pm 0.058_{stat} \pm 0.047_{syst}) \times 10^{-3}.
   \end{equation}
 At present most accurate measurement  of $K_S$ charge asymmetry was conducted by KLOE experiment.
	Measurement of $A_S$ was performed with $0.41 \mbox{ fb}^{-1}$ total
 luminosity data sample and the result is~\cite{kloe_final_semileptonic}:
  \begin{equation}
   A_{S} = (1.5 \pm 9.6_{stat} \pm 2.9_{syst}) \times 10^{-3}.
  \end{equation}
 Obtained charge asymmetry  for $K_{S}$ decays is consistent in error limits with charge asymmetry
	for $K_{L}$ decays.
 However, the inaccuracy of $K_S$ determination is more
 than two orders of magnitude bigger than this of the $A_L$ and the error of $A_S$ is 
 dominated by a statistical
 uncertainty, which is three times larger then systematical one. Therefore, in this work a new
	measurement of $A_S$ based on around four times bigger data sample collected by means of the KLOE
	detector	in 2004 and 2005, is presented.
	
\section{The K LOng Experiment at DA$\Phi$NE}
\begin{figure}[h!]
		\centering
		\includegraphics[width=0.6\textwidth]{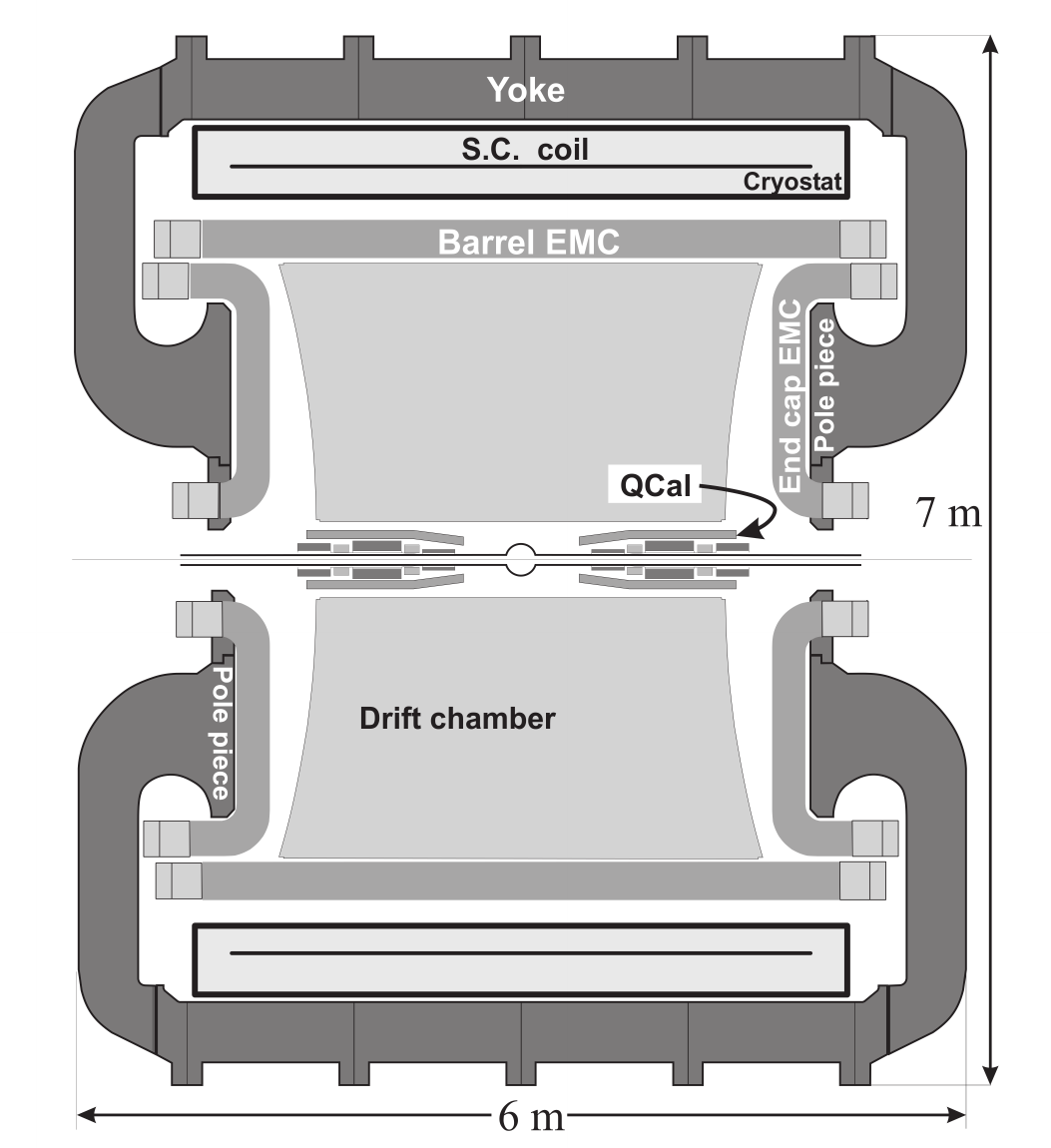}
		\caption{Scheme of the KLOE detector system. Drift chamber in the central part of the
		KLOE	detector is surrounded by the electromagnetic calorimeter. Both detectors	are inserted in a magnetic field.
		Figure adapted from \cite{KLOE_grey}.
		}
		\label{fig_kloe}
	\end{figure}

	KLOE was mounted at DA$\Phi$NE collider in 1999 
	and collected data during	two campaigns in 2001-2002 and 2004-2005. The gathered data sample corresponds	to the total
	luminosity of $2.5 \ \mbox{fb}^{-1}$. Energy of colliding beams ($e^-$ and $e^+$) was set to  
	the mass of $\phi$ meson. DA$\Phi$NE collider produce $\sim
	1300$ kaon pairs	per second which corresponds to the 	luminosity	$5 \times 10^{32} \ \mbox{cm}^{-2} \mbox{s}^{-1}$.
 In order to obtain efficient detection of the $K_L$ mesons the KLOE detector was constructed
 with a view to properties of neutral kaon system. 
	A schematic cross-section side view
	of	the KLOE detector is shown in Figure~\ref{fig_kloe}. The main components of KLOE are
	the	cylindrical drift chamber and the calorimeter, both surrounding the beam pipe.
 All elements are 	immersed	in a $0.52\ T$ magnetic field created by superconducting coils that are
	placed 	along	the beam axis.

	The	KLOE drift chamber (DC) is a $3.3\ \mbox{m}$ long cylinder with an internal and external radii of
	$25\ \mbox{cm}$	and $2\ \mbox{m}$, respectively, which  allows to register about $40\%$ of $K_L$ decays inside the chamber while
	the	rest reach the electromagnetic calorimeter. 
	Between the endplates around 12500 sense wires are streched and  allows
 to obtain: a spatial resolution in the $r, \varphi$ plane better than
 $200  ~\mu \mbox{m}$, a resolution along the $z$ axis of $\sim 2 \mbox{ mm}$ and  on the
	resolution of the decay vertex determination
 of  $\sim 1 \mbox{ mm}$.  Moreover, the curvature of the reconstructed tracks allows 
 to determine the particle
 momentum with a relative accuracy~of~$0.4~\%$. 
 The drift chamber allows to reconstruct information about charged particles while the calorimeter
 enables recording of both charged and neutral particles.
 The KLOE calorimeter has been designed to have  an excellent accuracy of energy determination 
 and  time resolution:
 \begin{align}
 \begin{aligned}
  \frac{ \sigma(E) }{ E} & =\frac{ 5.4 \%}{ \sqrt{E [\mbox{GeV}]}} \\
  \sigma_{t} & = \frac{54 \mbox{ ps}}{\sqrt{E [\mbox{GeV}] }} \oplus 140 \mbox{ ps}
 \end{aligned}
 \end{align}
 in order to register the hits of neutral particles and provide a possibility of the Time of Flight  (TOF)
 measurement.

\section{Registration of the $\phi \rightarrow K_L K_S \rightarrow K_L \pi e \nu$ processes at KLOE}
\subsection{Preselection}
\begin{figure}
		\includegraphics[width=0.45\textwidth]{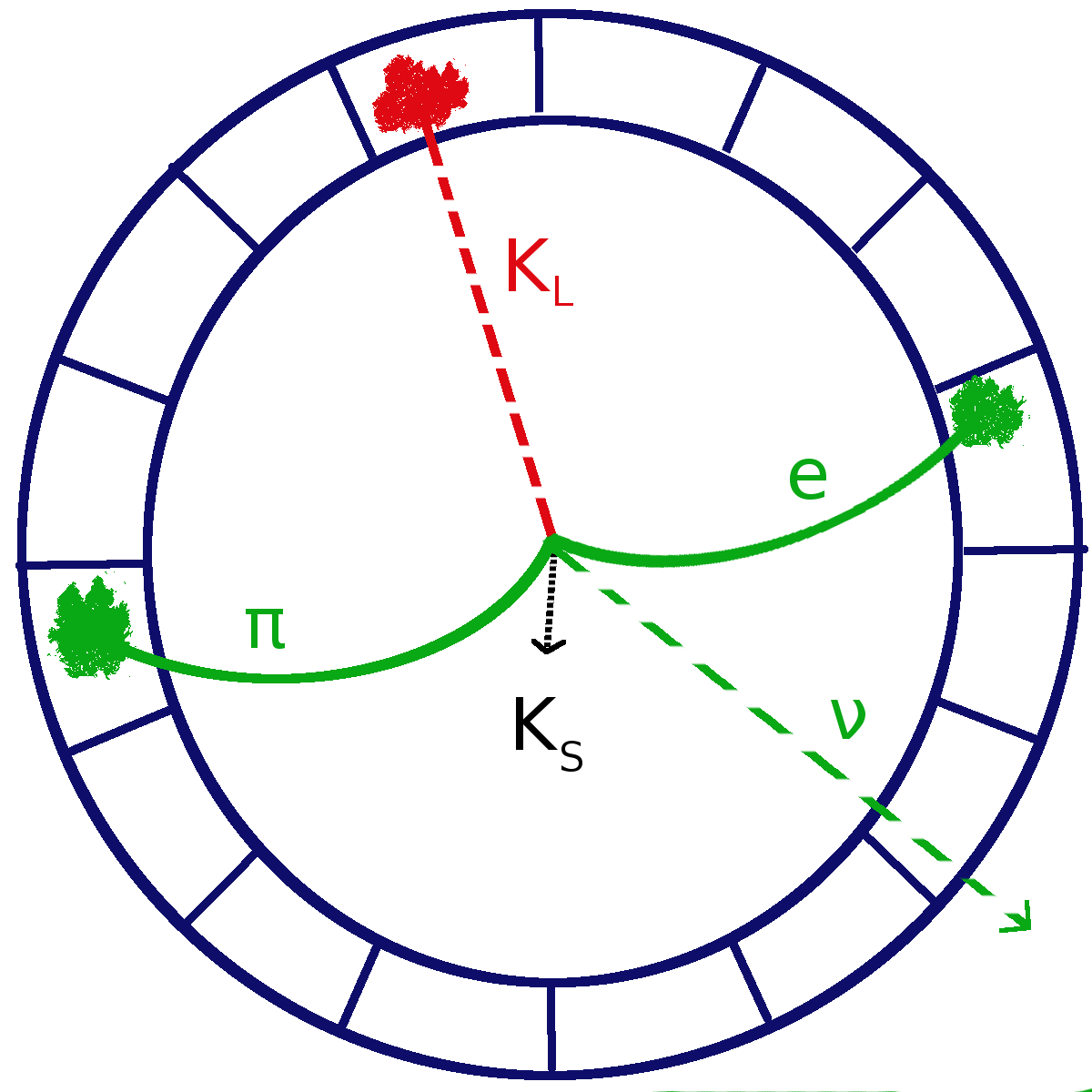}
		\includegraphics[width=0.45\textwidth]{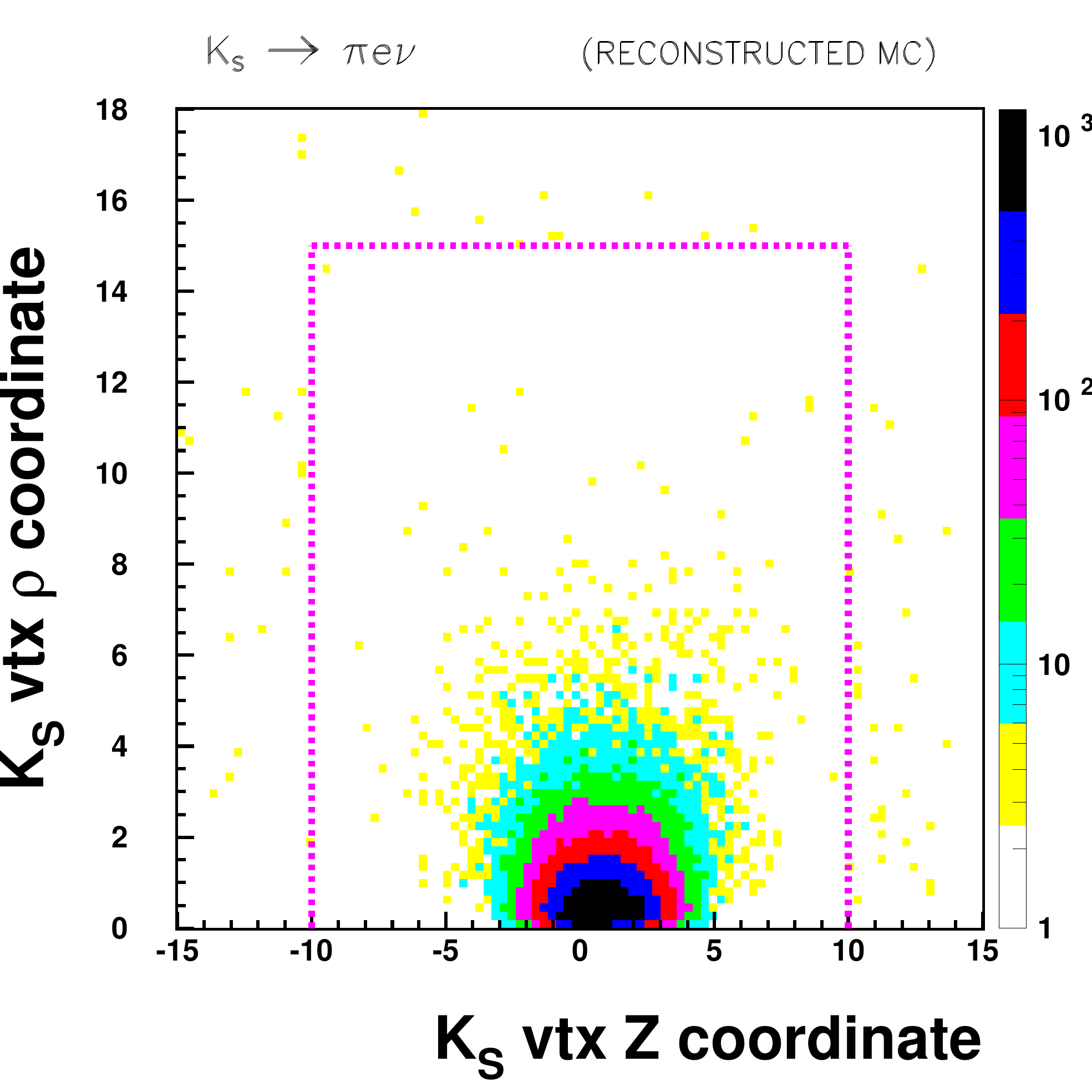}
  \caption{Left: Transverse view of exemplary signal event. The $K_{S}$ is identified using the $K_{L}$
  interaction in the calorimeter. In the next step  two oppositely charged
  particle tracks are selected which form a 
		vertex near the interaction point. Both tracks 
  must be associated with the colorimeter clusters. Right: Monte Carlo simulation
  of transversal and longitudinal coordinates of vertex position for  $K_{S} \rightarrow \pi e \nu$ events. Dashed
  line represents applied cuts which preserves $\sim 95 \%$ of the signal.}
		\label{events_sketch}
	\end{figure}

 Due to the conservation of quantum numbers during decay of $\phi$ meson the neutral kaons are
	produced in pairs. In general case it is an entangled anti-symmetric state, which properties allows
	to perform a fundamental tests of Quantum Mechanics and Lorentz symmetry~\cite{prospects_kloe}:
 \begin{equation}
  \ket{\phi}   = \frac{ \sqrt{(1+|\epsilon_{S}|^2)(1+|\epsilon_{L}|^2)}}{ \sqrt{2}(1- \epsilon_{S} \epsilon_{L})}
   \left( \ket{K_{L}( \vec{p})} \ket{K_{S}(-\vec{p}) }   
   -  \ket{K_{S}( \vec{p})} \ket{K_{L}(-\vec{p}) }  \right)   
		\label{phi_decay_into_neutral_kaons}	
 \end{equation} 
 where $\vec{p}$ is a momentum in the $\phi$ meson rest frame.
	However, when one of the kaon is
	detected at time $t >> \tau_{s}$, the state in Eq.~\ref{phi_decay_into_neutral_kaons} factorizes
	and the system behaves as if the initial state was a mixture of $\ket{K_{L}( \vec{p})}
	\ket{K_{S}(-\vec{p})}$ and $\ket{K_{S}( \vec{p})} \ket{K_{L}(-\vec{p})}$. Hence the detection of
	$K_L$ at large distance from Interaction Point tags $K_S$ state in the opposite direction.
	Measurement described in this paper is based on identification of $K_S$ through the detection of
	$K_L$ interaction in the calorimeter. 
 A sketch of typical signal event is shown in Figure~\ref{events_sketch}(left).
 Selection of $K_L$ candidates takes into account only
	clusters with high energy deposition and requires that the cluster is not close to any track
	reconstructed in the drift chamber. Also, velocity of neutral particle  that deposits energy in
	this cluster must be close to the theoretical value of $K_L$ meson velocity in the $\phi$ meson rest
	frame ($\beta \sim 0.22$).	 	
	
\begin{figure}[h!]
		\includegraphics[width=0.45\textwidth]{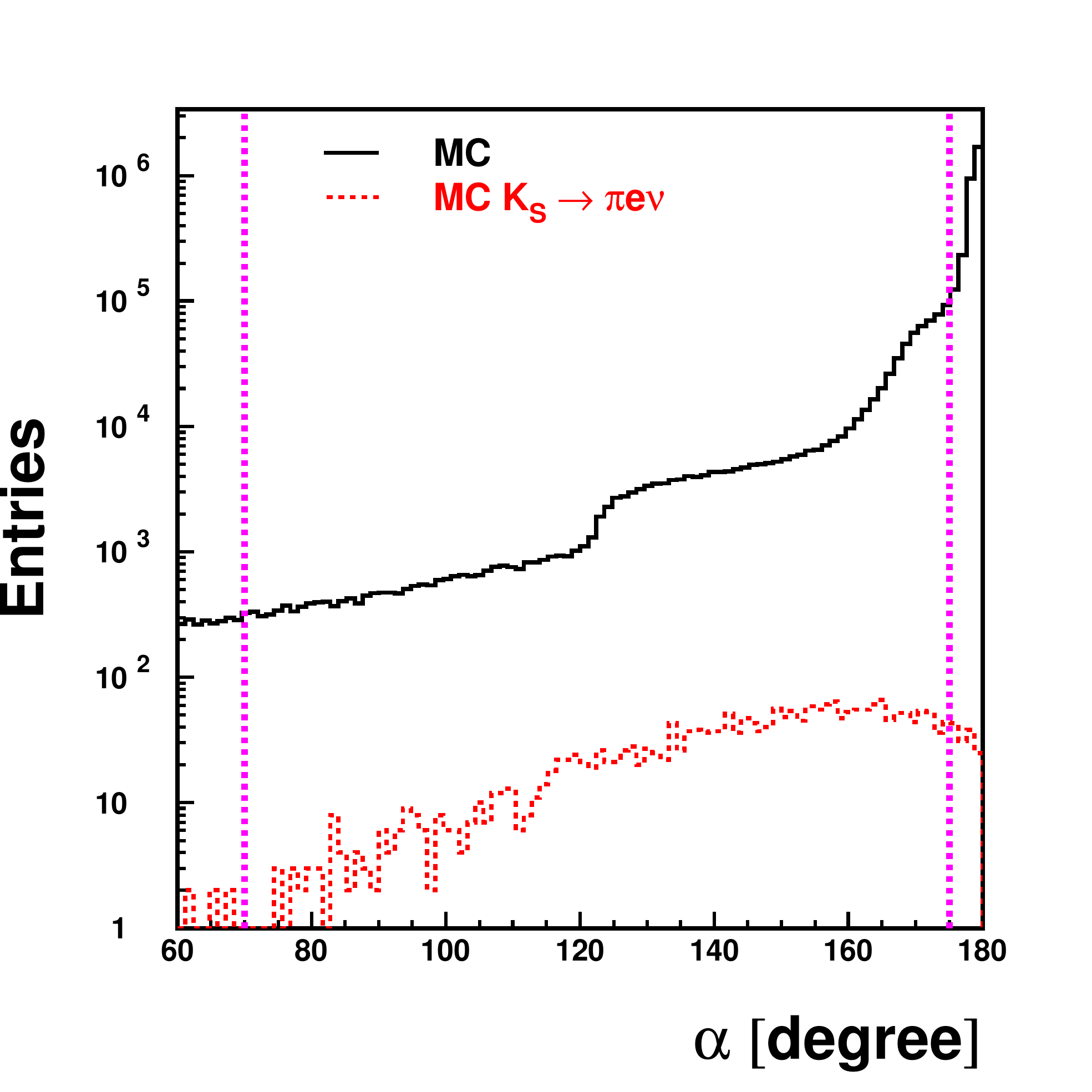}
		\includegraphics[width=0.45\textwidth]{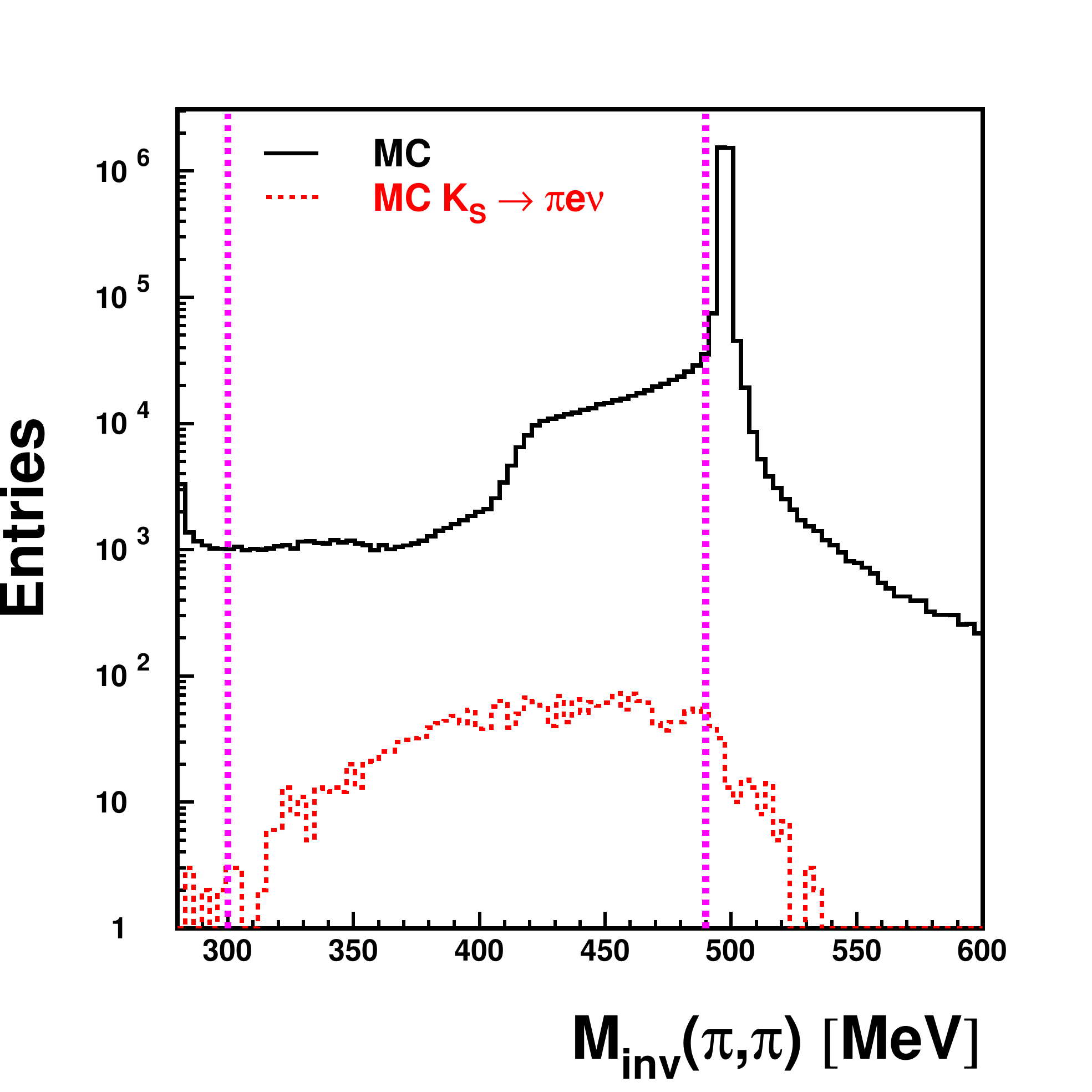}
		\caption{Left: Simulated distribution of angle between charged secondaries in $K_S$ rest
  frame. Right: Simulated distribution of invariant mass calculated under assumption that both registered particles were pions.
  In both figures solid and dashed histograms represents all events and semileptonic decays, respectively.
  Vertical dashed lines represent cuts described in text.}
		\label{cuts_angle_invmass}
	\end{figure}

	In the next step, candidates for $K_S$ meson decays are selected by selection of two oppositely
	charged particles with	tracks	forming  a vertex close to the interaction point~(IP):
		\begin{equation}
			\begin{aligned}
				\rho_{vtx}	< 15 \mbox{ cm}, \\
				|z_{vtx}|	< 10 \mbox{ cm},
			\end{aligned}
		\end{equation}
	where	$\rho_{vtx} = \sqrt{ x_{vtx}^{2} + y_{vtx}^{2}} $. Obtained distribution is shown in
	Figure~\ref{events_sketch}(right). Then, the rejection of main source of background ($K_S \rightarrow
	\pi^+ \pi^-$) is conducted by applying the following cuts:
		\begin{itemize}
   \item $70^{\circ}<\alpha <175^{\circ}$ \\ where $\alpha$ is an angle between charged secondaries in $K_{S}$ rest frame.
    Obtained $\alpha$ distribution is shown at the left side of Figure~\ref{cuts_angle_invmass}.
    In case of three body decay ($K_{S} \rightarrow \pi e \nu$) it is spanned over a large angle range.
   \item $300 <M_{inv}< 490 \mbox{ MeV}$ \\ 
   an invariant mass $M_{inv}$ is calculated using  momenta of the particles which tracks form a~vertex
   assuming that both particle were pions.
   Obtained $M_{inv}$ distribution is shown at the right side of Figure~\ref{cuts_angle_invmass}.
		\end{itemize}
 Both tracks reconstructed in the drift chamber must be associated with neighbouring
 clusters in calorimeter in order to use Time-of-Flight technique.

\subsection{$K_S \rightarrow \pi e \nu$ events identification}
 \begin{figure}
  \centering
  \includegraphics[width=0.49\textwidth]{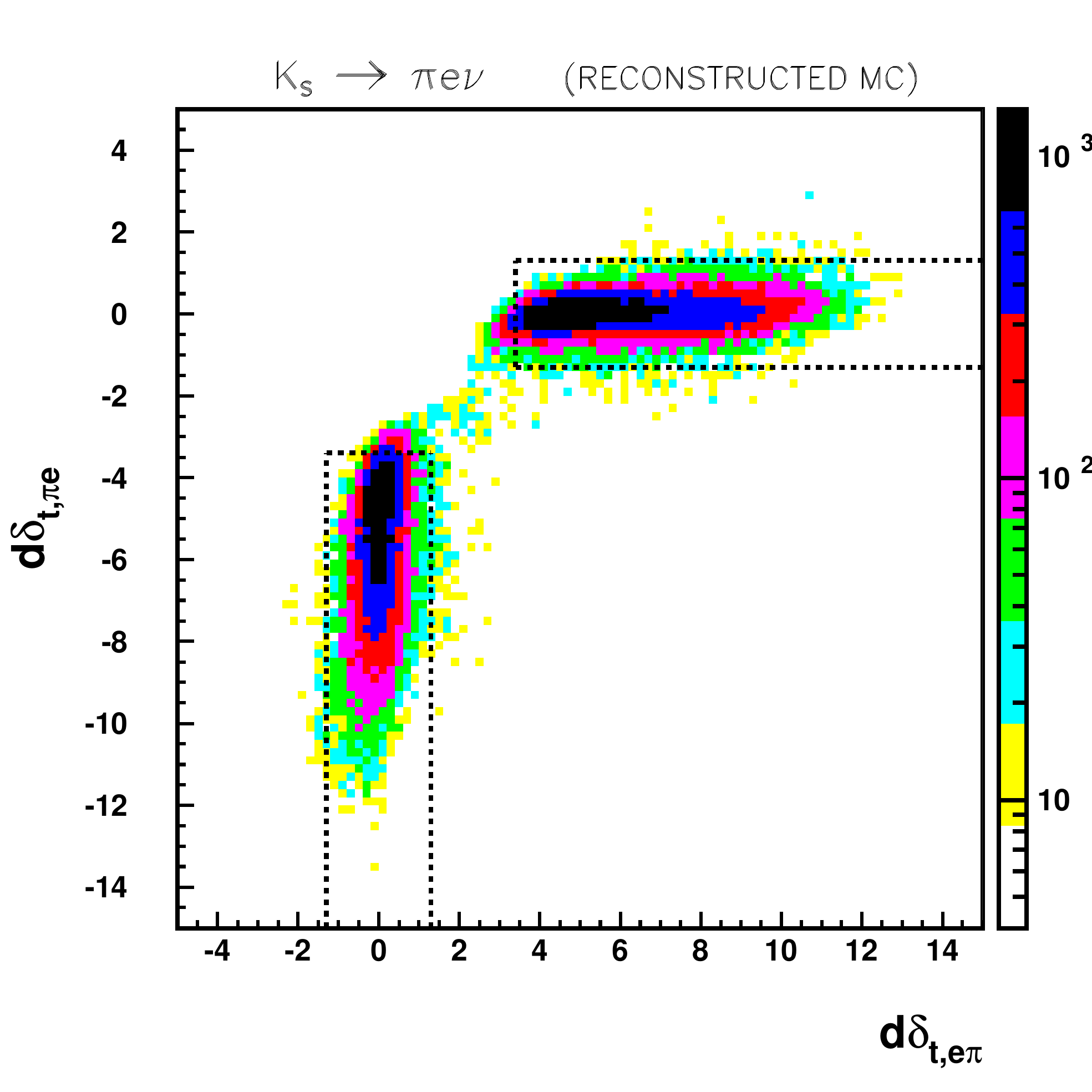}
  \includegraphics[width=0.49\textwidth]{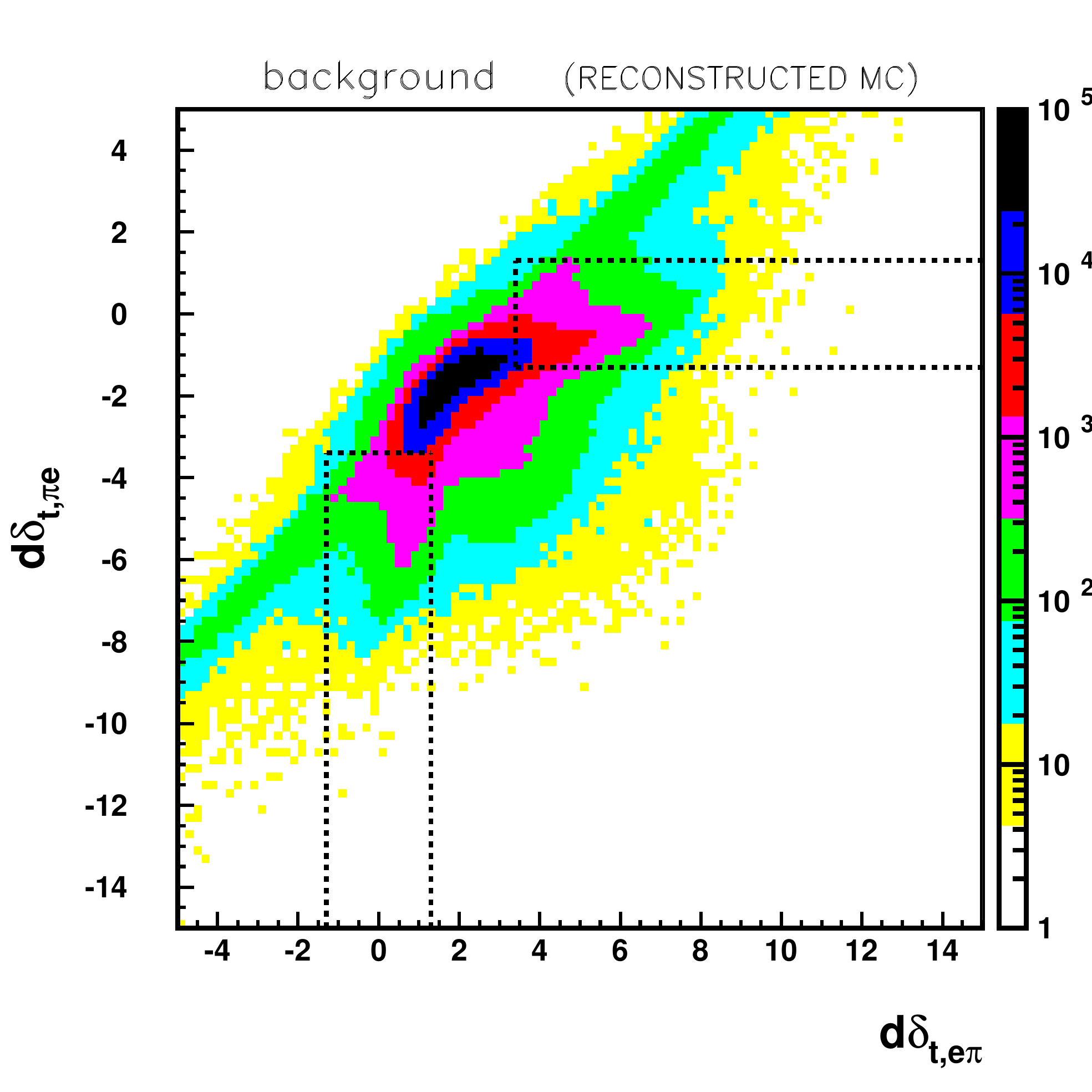}
  \caption{Simulated distributions of time differences $d \delta_{t, \pi e}$ vs $d \delta_{t, e \pi}$ defined
  in Eq.~\ref{tof2d_mpi} for $K_{S}\rightarrow \pi e \nu$ events (left) and background events
  (right), once the $\delta_{t,\pi \pi}$ cut has been applied. The regions delimits by the dashed
  lines are selected. 
  In case of $e\pi$ ($\pi e$)  the $d \delta_{t, e \pi}$ ($d \delta_{t, \pi e}$)
  variable  acquires  value close to zero. 
  }
  \label{rys_tof2D}
 \end{figure}

 \begin{figure}
  \includegraphics[width=0.49\textwidth]{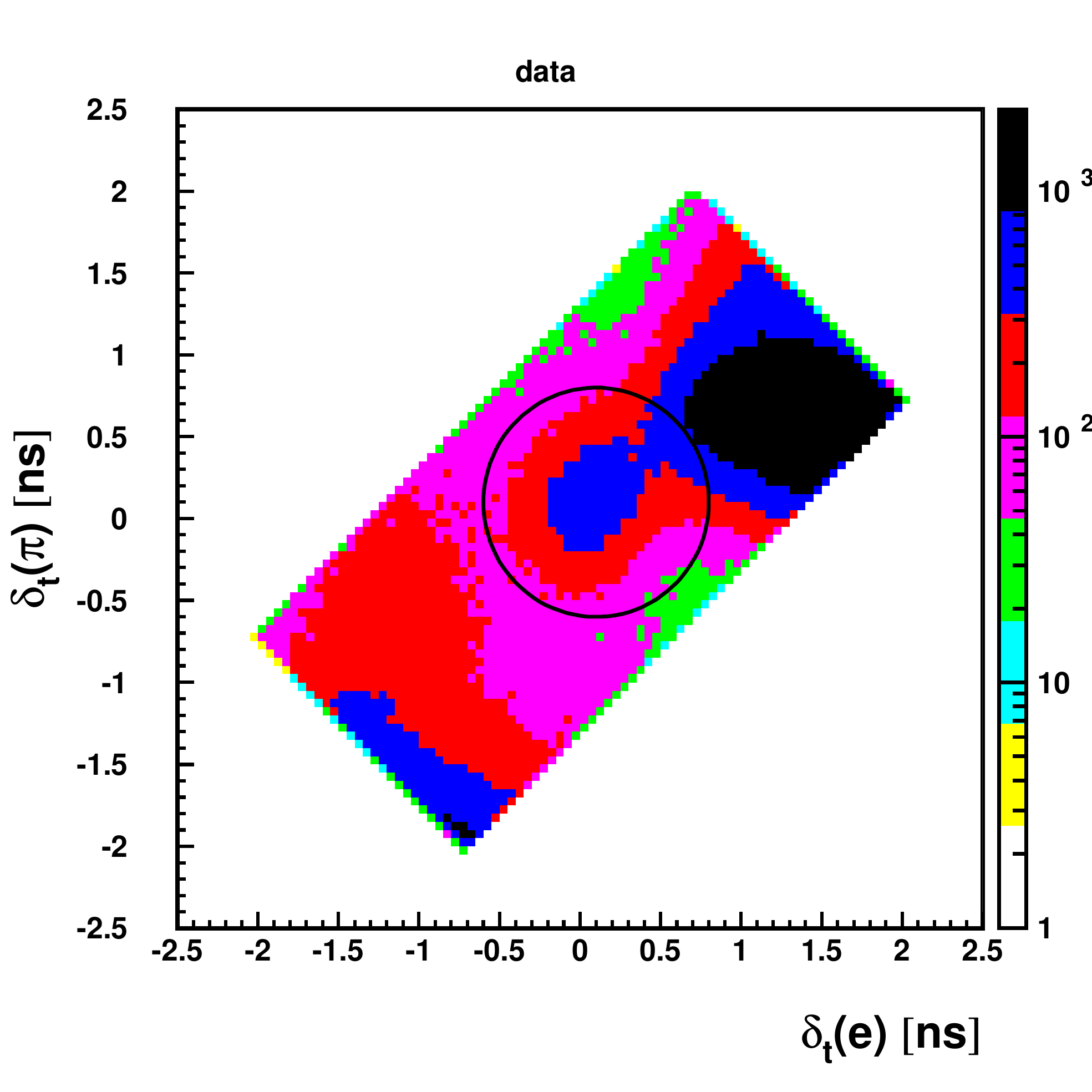}
  \includegraphics[width=0.49\textwidth]{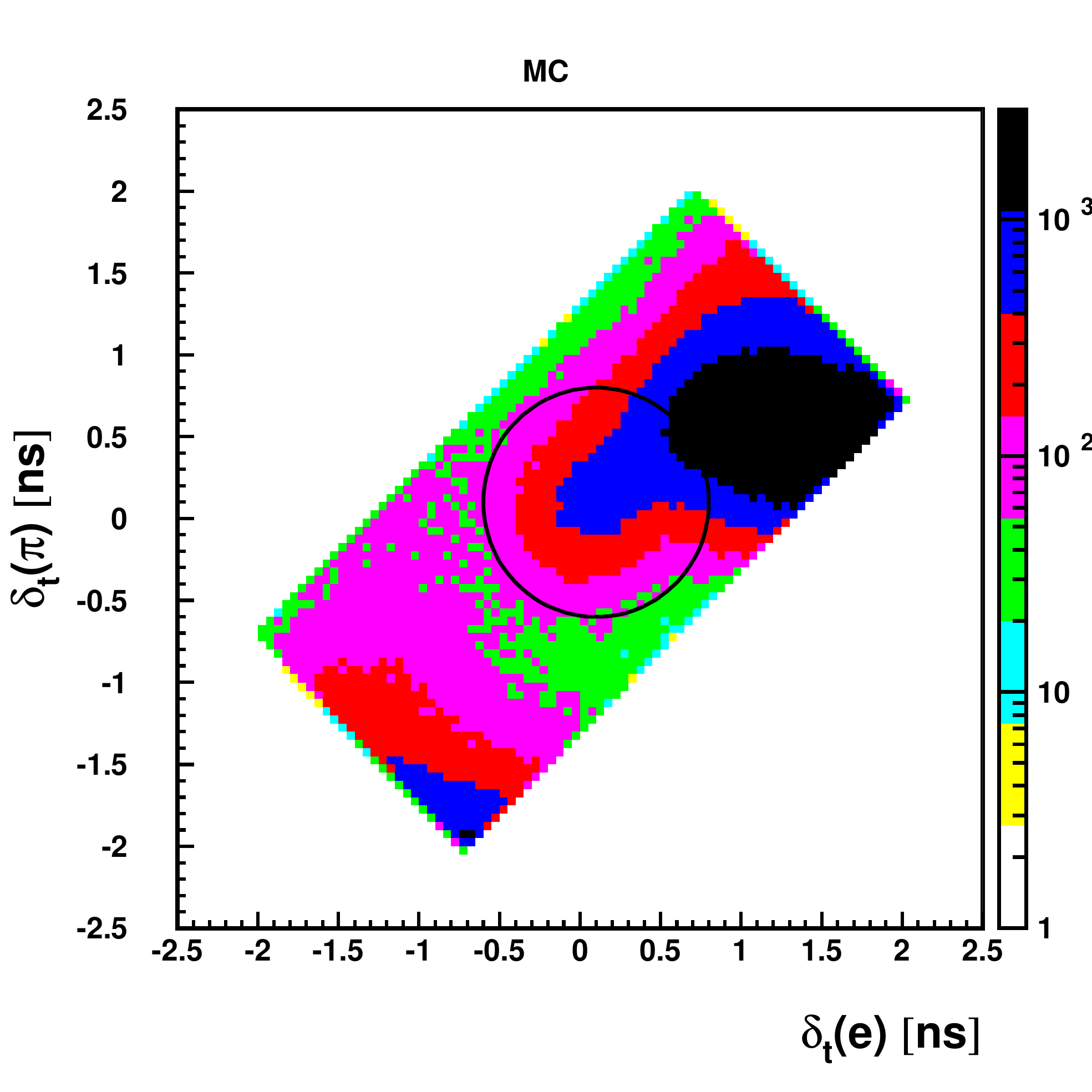}
  \newline
  \includegraphics[width=0.49\textwidth]{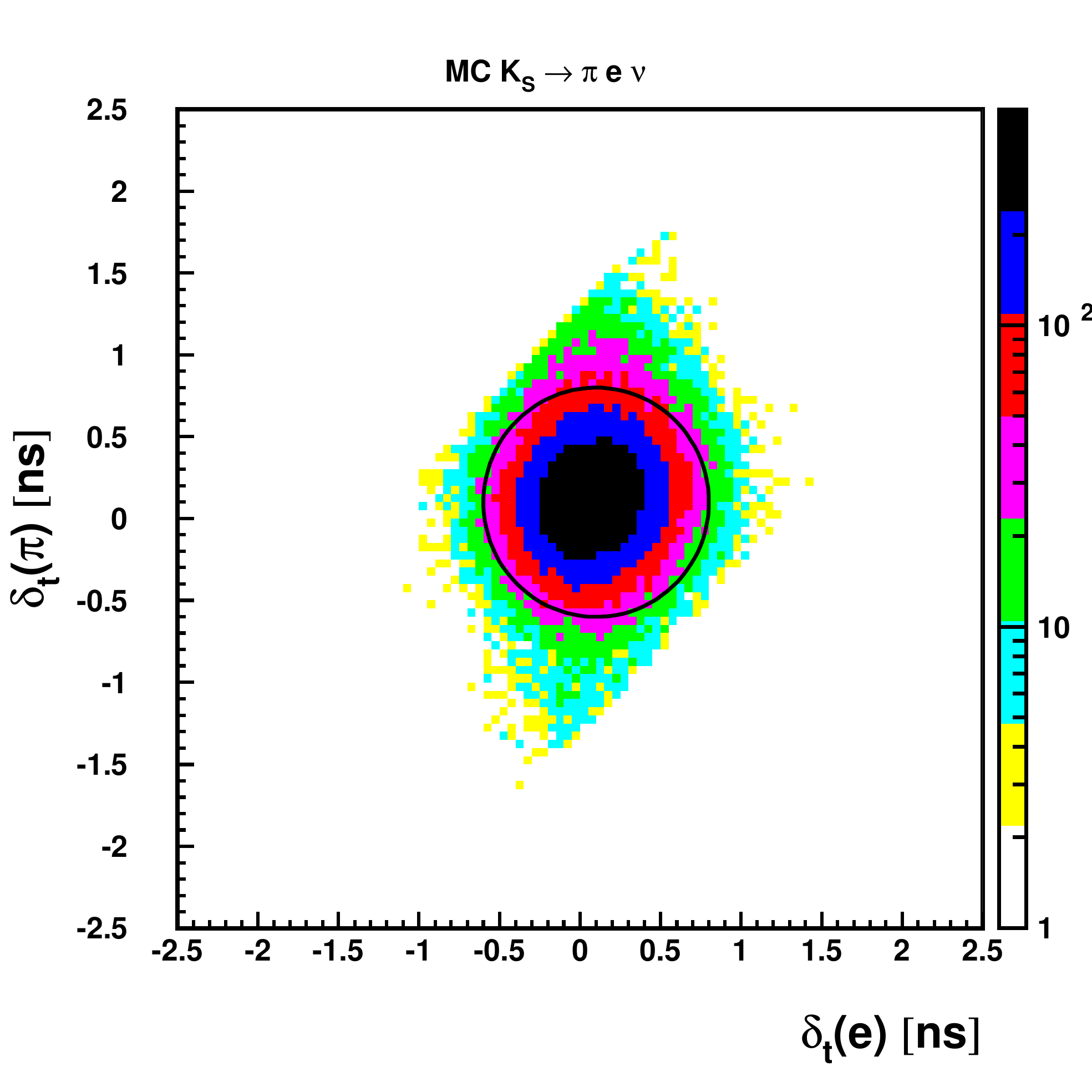}
  \includegraphics[width=0.49\textwidth]{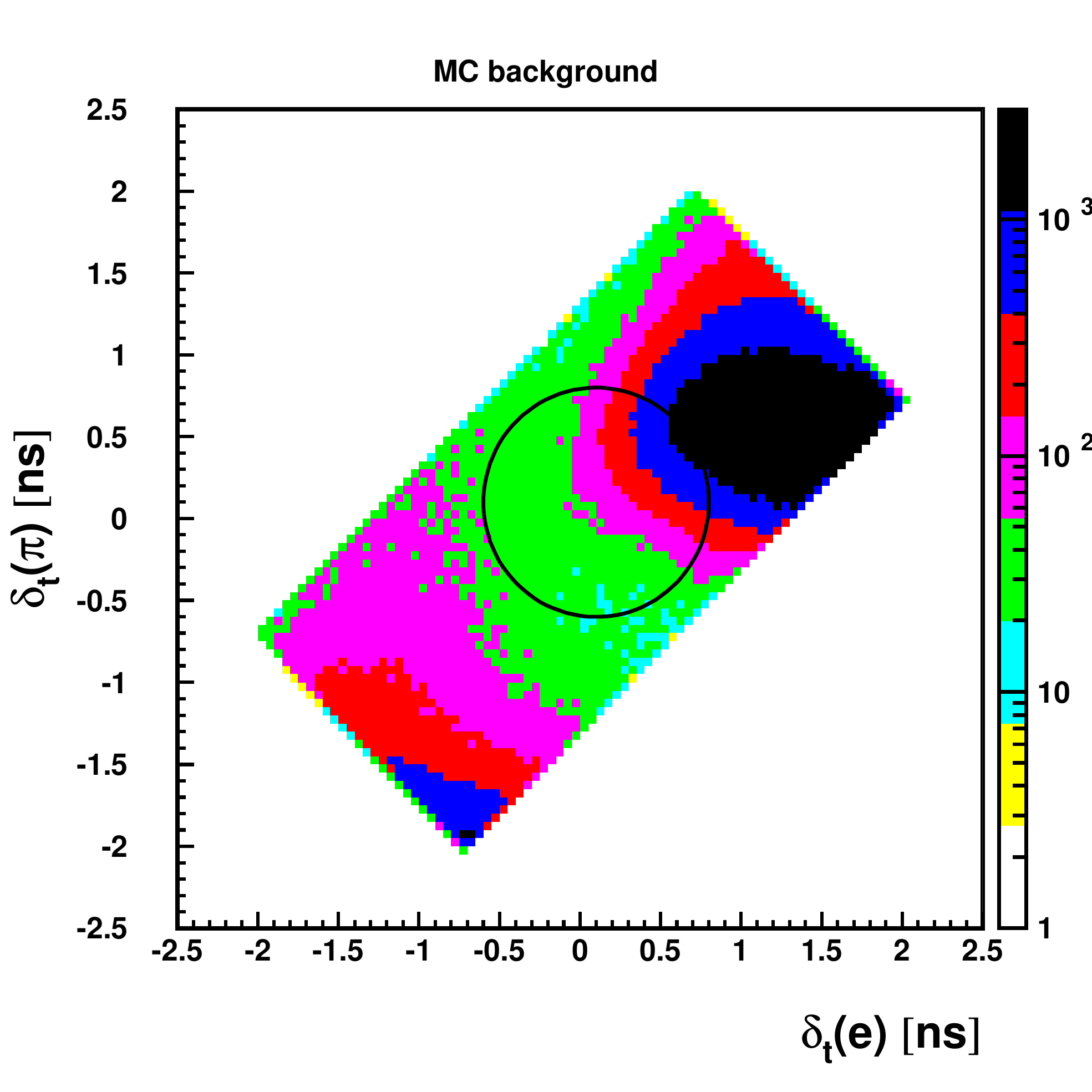}
  \caption{Distributions of the time difference for pion mass hypothesis ($\delta_{t}(\pi)$) versus
  the time difference for electron mass hypothesis ($\delta_{t}(e)$) for experimental data
  (top-left), total MC events (top-right), MC $K_{S} \rightarrow \pi e \nu$ events (bottom-left)
  and MC background events (bottom-right). Events within the circle 
  are retained for further analysis.
  }
  \label{recalculated_dt}
 \end{figure}

 The Time of Flight technique aims at rejection of the background, which at this stage of analysis is
 due to $K_{S} \rightarrow \pi^+ \pi^-$ events, and  at identification of the final charge
 states ($\pi^{\pm}  e^{\mp}$).   
 For each particle, the difference $\delta_{t}$ between the measured time of associated cluster 
 ($t_{cl}$) and expected time of flight is calculated assuming a given mass hypothesis,~$m_{x}$:
  \begin{equation}
   \begin{aligned}
     \delta_{t}(m_{x}) = t_{cl} - \frac{L}{c \cdot \beta(m_{x})},  \\
     \beta(m_{x}) =  \frac{P}{ \sqrt{P^2 + m_{x}^2 }},
   \end{aligned}
  \end{equation}
 where $L$ is a total length of particle trajectory  and $P$ is particle momentum.
 For further consideration it is useful to introduce the
  difference of $\delta_{t}(m_a)$ and $\delta_t(m_b)$: 
  \begin{equation}
   d \delta_{t,ab} = \delta_{t}(m_{a})_{1} - \delta_{t}(m_{b})_{2}.
  \end{equation}
	where suffix 1 and 2 refers to the first and second particle according to the ordering determined
	by the reconstruction procedure.	

 Two cuts are further applied. For the first cut, both particles are assumed to be pions and
 $d \delta_{t,\pi \pi}$  is calculated. For $K_{S} \rightarrow \pi \pi$ events this value is around zero and this fraction of events could be
 rejected by requiring:
  \begin{equation}
   | d \delta_{t, \pi \pi}| > 1.5 \mbox{ ns}.
  \end{equation}
	\begin{figure}
  \centering
	 \includegraphics[width=0.6\textwidth]{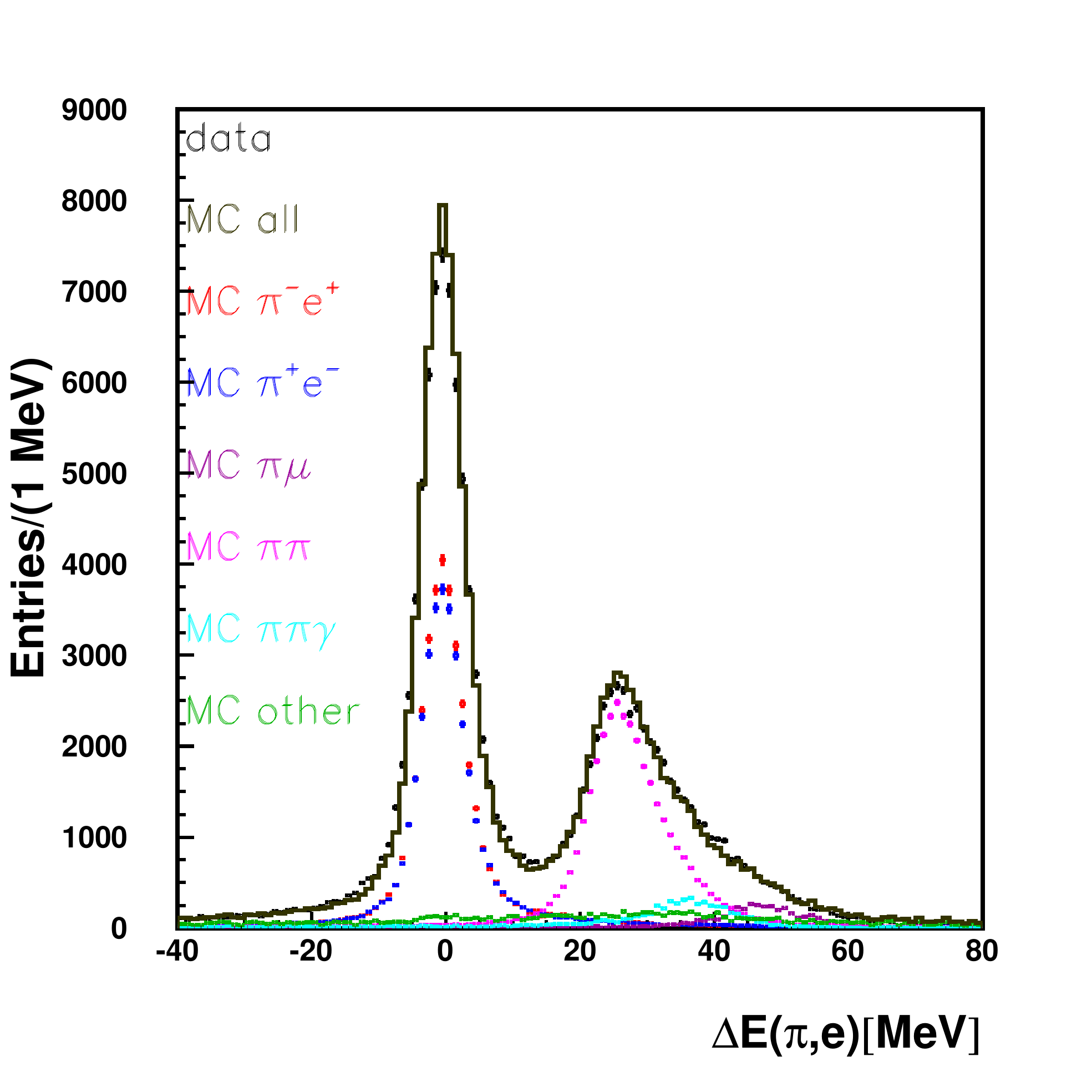}
	 \caption{Distribution of $\Delta E(\pi,e) = E_{miss} - p_{miss}$ for all selected events 
	 after normalization procedure.}
	 \label{de_pi_e}
	\end{figure}

 Then, for surviving events, the pion-electron hypothesis is tested:
  \begin{equation}
   \begin{aligned}
   d \delta_{t, \pi e} = \delta_t(m_{\pi})_{1} - \delta_{t}(m_{e})_{2}, \\
   d \delta_{t, e \pi} = \delta_t(m_{e})_{1} - \delta_{t}(m_{\pi})_{2}.
   \label{tof2d_mpi}
   \end{aligned}
  \end{equation}
 If a mass assumption is correct then one of the variables above should be close to zero. The
 obtained  Monte Carlo distributions for $K_{S} \rightarrow \pi^{\pm} e^{\mp} \nu$ and
 background events are presented in Figure~\ref{rys_tof2D}.
 Hence, the following cut is applied:
 \begin{equation}
  \begin{aligned}
   |d \delta_{t, e \pi}| < 1.3 \mbox{ ns} ~\wedge~ d \delta_{t,  \pi e} < -3.4 \mbox{ ns} \\
     \mbox{or ~~~~~~~~~~~~~~~~~~~~~} \\
   d \delta_{t, e \pi} > 3.4 \mbox{ ns} ~\wedge~ |d \delta_{t,  \pi e} |< 1.3 \mbox{ ns}
  \end{aligned}
 \end{equation} 
 The above requirement ensures that the possibility of misidentification of charged particles from
 signal events is equal to $10^{-4}$ only.

 Obtained distributions for simulated data and experimental KLOE data are shown in Figure~\ref{recalculated_dt} (top).
 The Monte Carlo simulation indicates the signal position around the point (0,0) while the
 background location is spread at the corners of the obtained distribution. 
 Due to that an additional TOF cut is applied by selecting events within the circle  
 in the  $\delta_{t}(e)$ vs $\delta_{t}(\pi)$ plane, as shown in  Figure~\ref{recalculated_dt}.  
 This cut allows to preserve $94\%$ of the remaining signal and control the number of 
 selected background events for normalization. 
 The distribution of
 the difference between the missing energy and momentum ($\Delta E(\pi,e)$) shows
 the remaining background components (see Figure~\ref{de_pi_e}). Based on an integrated
 luminosity of $1.7 \mbox{fb}^{-1}$ around $10^5$ of $K_S \rightarrow \pi e \nu$ decays were reconstructed
 and will be used to determine the charge asymmetry and branching ratio
 for $K_S$ semileptonic decays. A preliminary analysis shows a potential of
 reaching a two times better statistical error determination with a sample
 four times bigger than the previous KLOE analysis. The analysis is still in
 progress and preliminary results will be available soon.

\section{KLOE-2 Project}
	DA$\Phi$NE collider in previous years was adapted to increased delivered luminosity by installation
	of the new interaction region based on the 	Crabbed Waist compensation together with large Piwinski
	angle~\cite{dafne_upgrade}. Those changes should increase by factor of three the amount of the delivered luminosity with
	respect to the performance reached before.
	Together with  the upgrade of the KLOE detector this will allow to collect by KLOE-2 project the
	order of $10\mbox{ fb}^{-1}$ of integrated luminosity.
	The	detector itself was equipped with crystal  (CCALT)~\cite{CCALT} and tile (QCALT)~\cite{QCALT} calorimeters
	which  covers the low polar angles 	and  improve the detection of the photons that are coming  from
	$K_L$ decays in the drift chamber, respectively. Also, there were mounted the two pairs of small
	angle tagging devices that allows to detect the low (Low Energy Tagger~\cite{LET}) and high
	(High Energy Tagger~\cite{HET}) energy $e^+ e^-$ originated from $e^+ e^- \rightarrow e^+ e^-
	X$ reactions. However, the especially important for studies of $\mathcal{CPT}$ symmetry violation
	through semileptonic decays
	in neutral kaons is precise reconstruction of tracks momenta and vertex reconstruction  close to
	the Interaction Point. This	will be provided by Inner Tracker detector made out in a novel GEM technology~\cite{inner_tracker}.  
 Those upgrades will also improve the sensitivity of $\mathcal{CPT}$ and Lorentz invariance tests,
	which were presented  by the KLOE experiment and became the  most precise measurement of the $\mathcal{CPT}$
	violating parameters $\Delta a_{\mu}$ for neutral kaons in the Standard Model Extension
	(SME)~\cite{cpt_antonio}. 
	Measured values of $\Delta a_{\mu}$ currently have a precision of $10^{-18}$ GeV and are proportional to the
	parameter $\delta_{K}$~\cite{kostelecky_1,kostelecky_2,kostelecky_3}. 
	

	It	should be emphasised that KLOE-2 aims to significantly
	improve	the sensitivity of tests of discrete symmetries, through studies of
	$K_S$ charge asymmetry or quantum interferometry effects in the kaon decays,
	beyond	the presently achieved results.

\ack
	We warmly thank our former KLOE colleagues for the access to the data collected during the KLOE data taking campaign.
 We thank the DA$\Phi$NE team for their efforts in maintaining low background running conditions and their collaboration during all data taking. We want to thank our technical staff: 
 G.F. Fortugno and F. Sborzacchi for their dedication in ensuring efficient operation of the KLOE computing facilities; 
 M. Anelli for his continuous attention to the gas system and detector safety; 
 A. Balla, M. Gatta, G. Corradi and G. Papalino for electronics maintenance; 
 M. Santoni, G. Paoluzzi and R. Rosellini for general detector support; 
 C. Piscitelli for his help during major maintenance periods. 
 This work was supported in part by the EU Integrated Infrastructure Initiative Hadron Physics
	Project under contract number RII3-CT- 2004-506078; by the European Commission under the 7th
	Framework Programme through the `Research Infrastructures' action of the `Capacities' Programme,
	Call: FP7-INFRASTRUCTURES-2008-1, Grant Agreement No. 227431; by the Polish National Science Centre
	through the Grants No. 
 DEC-2011/03/N/ST2/02641, 
 2011/01/D/ST2/00748,
 2011/03/N/ST2/02652,
 2013/08/M/ST2/0 0323,
	DEC-2014/12/S/ST2/00459,
	2013/11/B/ST2/04245
 and by the Foundation for Polish Science through the MPD programme and the project HOMING PLUS BIS/2011-4/3.


\section*{References}
\begin{thebibliography}{9}
	\bibitem{cpt_table} V. A. Kostelecky, N. Russell, Rev. Mod. Phys. {\bf 83} (2011), 11
 \bibitem{cpt_lorentz} O. W. Greenberg, Phys. Rev. Lett. {\bf 89} (2002), 231602
 \bibitem{handbook_cp} L. Maiani, G. Pancheri, N. Paver,  INFN-LNF (1995)
 \bibitem{ktev_kl_charge_asymm} A. Alavi-Harati et al. (KTeV Collaboration),  Phys. Rev. Lett. {\bf 88} (2002), 181601
 \bibitem{kloe_final_semileptonic} F. Ambrosino et al. (KLOE Collaboration), Phys. Lett. {\bf B636}
 (2006), 173
	\bibitem{KLOE_grey} G. Vignola, M. Bassetti, M. E. Biagini, C. Biscari, R. Boni, 
	Conf. Proc. {\bf C960610} (1996), 22
 \bibitem{prospects_kloe} G. Amelino-Camelia et al., Eur. Phys. J. {\bf C68} (2010), 619
	%
 \bibitem{drift_chamber} Adinolfi, M. et al., (KLOE Collaboration), Nucl. Instrum. Meth. {\bf A461} (2001), 25
 \bibitem{calorimeter}  Adinolfi, M. et al., (KLOE Collaboration), Nucl. Instrum. Meth. {\bf A482} (2002), 364
 \bibitem{dafne_upgrade} Milardi, C. er al., (DA$\Phi$NE Collaboration), CERN-2008-006 (2008), 1 
 \bibitem{CCALT} M. Cordelli et al., Nucl. Instrum. Meth. {\bf A718} (2013),  81
 \bibitem{QCALT} A. Balla et al., Nucl. Instrum. Meth. {\bf A718} (2013), 95
 \bibitem{LET}  M. Adinolfi et al., (KLOE Collaboration), Nucl. Instrum. Meth. {\bf A617} (2010), 81
 \bibitem{HET}  M. Adinolfi et al., (KLOE Collaboration), Nucl. Instrum. Meth. {\bf A617} (2010), 266
 \bibitem{inner_tracker} G. Morello et al., JINST {\bf 9} (2013),  C01014
	\bibitem{cpt_antonio} D. Babusci et al., (KLOE Collaboration), Phys. Lett {\bf B730} (2014), 89
	\bibitem{kostelecky_1} D. Colladay and V. A. Kostelecky, Phys. Rev. {\bf D55}, (1997) 6760
	\bibitem{kostelecky_2} D. Colladay and V. A. Kostelecky, Phys. Rev. {\bf D58}, (1998) 116002
	\bibitem{kostelecky_3} V. A. Kostelecky and R. Potting, Phys. Rev. {\bf D51}, (1995) 3923
\end{thebibliography}
\end{document}